\documentstyle[times,epsfig]{jaa}
\begin{document}
\title[]{A Non-Mainstream Viewpoint on Apparent Superluminal Phenomena in AGN Jet}

\author[Wen-Po Liu {\it et al.}]%
       {Wen-Po Liu$^1$\thanks{Correspondence author: wp-liu@cauc.edu.cn}, Li-Yan Liu$^1$ $\&$ Chun-Cheng Wang$^2$ \\
$^1$College of Science, Civil Aviation University of
China, Tianjin 300300, China \\
$^2$Key Laboratory for Research in Galaxies and Cosmology, University of Science \\
and Technology of China, Chinese Academy of Sciences, Hefei, Anhui 230026, China \\}
%\pubyear{}
%\volume{}
%\setcounter{page}{}
\date{}
\maketitle
\label{firstpage}
\begin{abstract}
The group velocity of light in material around the AGN jet
is acquiescently one (c as a unit), but this is only a hypothesis. Here,
we re-derive apparent superluminal and Doppler formulas for the general
case (it is assumed that the group velocity of light in the uniform and
isotropic medium around a jet (a beaming model) is not necessarily equal
to one, e.g., Araudo et al. (2010) thought that there may be dense clouds
around AGN jet base), and show that the group velocity of light close to
one could seriously affect apparent superluminal phenomena and Doppler
effect in the AGN jet (when the viewing angle and Lorentz factor take
some appropriate values).

\end{abstract}

\section{Introduction}
\label{sec:intro}
Concerning the apparent superlunimal motion, the most popular and
classic viewpoint is the relativistic beaming model (Rees 1966). The
Fig. S.6 in Rybicki \& Lightman (2004) showed the simple geometry
scenario of emission for a moving source, the observed transverse
velocity of separation of a blob relative to the speed of light $c$
(in this paper, all velocities are at $c$ as a unit)$\colon$
\begin{equation}
\beta_a=\frac{\beta sin \theta}{1-\beta cos \theta},
\end{equation}
\noindent where $\beta$ is the true velocity, and $\theta$ is the
angle to the line of sight.
   The corresponding Doppler factor $\delta$ (considering the effect of redshift) is$\colon$
\begin{equation}
\delta=[\Gamma (1+z) (1-\beta cos \theta)]^{-1},
\end{equation}
\noindent where $\Gamma=(1-\beta ^2)^{-1/2}$ is the Lorentz factor,
$z$ is the redshift of this AGN.

For the 3C 273 jet ($z=0.158$, Schmidt 1963), VLBI observations have detected
apparent superluminal motions in the parsec-scale jet with apparent
velocities 6 $\sim$ 10 (e.g., Unwin et al. 1985). We assume the
apparent velocity is 8, and the angle to the line of sight is
10$^\circ$. Then, based on the formula (1) and (2), we could obtain
the Doppler factor $\delta$ $\sim$ 4.5, and the true velocity $\sim$
0.994 (Lorentz factor $\Gamma$ $\sim$ 8.8) which means the velocity of
the `ordinary' matter in the AGN jet is extremely close to the speed
of light.

The formulas (1) $\&$ (2) actually imply that the group velocity of light in
the medium surrounding a blob in an AGN jet is equal to one, which is
actually a hypothesis. In the following, we will consider a general
scheme.

\begin{figure*}
\includegraphics[]{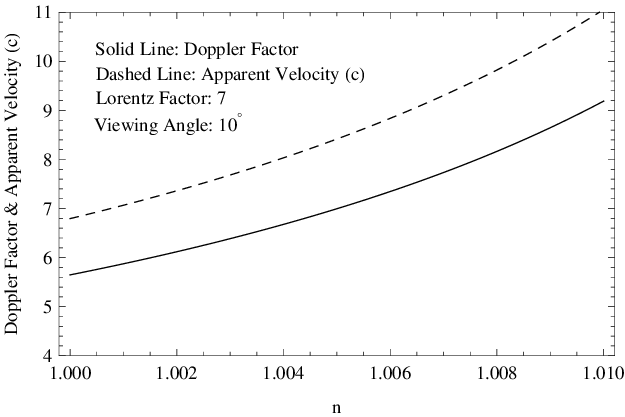}
\caption{ The plot shows that the apparent velocity and Doppler factor change
with $n$ from 1 to 1.01 (corresponding to the group velocity of light from 0.99 to 1).
We assume $z=0$, $\theta=10^\circ$ and $\Gamma=7$. }
\label{fig1}
\end{figure*}

\section{Model}
\label{sec:using}
We assume that the medium surrounding a blob is
uniform, transparent, isotropic£¬and stationary relative to the AGN core, and the
group velocity of light is $\beta_g$ which is not necessarily equal to one
($\beta_g$ may be a function of frequency).

Then we could apply the similar derivation like Rybicki \& Lightman
(2004) and get the `new' (modified) apparent velocity and Doppler
formula$\colon$
\begin{equation} \beta_a=\frac{\beta sin \theta}{1-n \beta cos
\theta},
\end{equation}

\begin{equation}
\delta=[\Gamma (1+z) (1-n \beta cos \theta)]^{-1},
\end{equation}
\noindent Where $n=1/\beta_g$ (if $\beta_g$ is equal to the phase velocity of light in material,
then $n$ means refractive index of material).

\begin{figure*}
\includegraphics[]{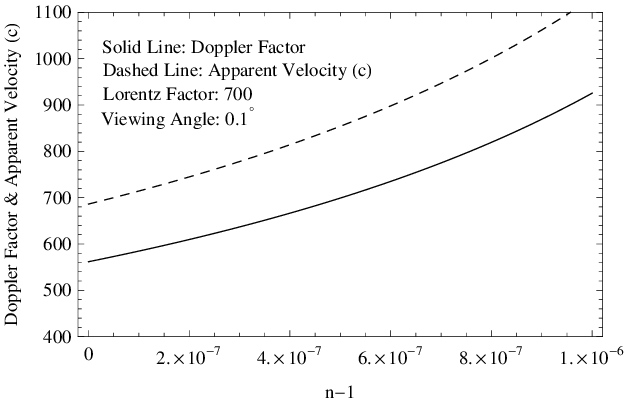}
\caption{ The plot shows that the apparent velocity and Doppler factor change
with $n$ from 1 to 1.000001 (corresponding to the group velocity of light from 0.999999 to 1).
We assume $z=0$, $\theta=0.1^\circ$ and $\Gamma=700$. }
\label{fig2}
\end{figure*}

\section{Discussion}
We apply the formulas (3) $\&$ (4) to the 3C 273 jet (we assume that
$\beta_a=8$, $\theta=10^\circ$, $n=1.01$ corresponding to $\beta_g=0.99$) and obtain the Doppler
factor $\delta$ $\sim$ $7.2$, the true velocity $\sim$ 0.984 (Lorentz factor $\Gamma$ $\sim$ 5.6)
which are clearly different from the ones in the case of $n=1$ corresponding to $\beta_g=1$.

Fig.~\ref{fig1} and Fig.~\ref{fig2} show that the apparent velocity and Doppler factor change
with $n$ (For Fig.~\ref{fig1}, we take $z=0$, $\theta=10^\circ$ and $\Gamma=7$;
In Fig.~\ref{fig2}, we assume that $z=0$, $\theta=0.1^\circ$ and $\Gamma=700$).

As shown, when the viewing angle and Lorentz factor take some appropriate values,
the group velocity of light close to one could still seriously affect apparent superluminal
phenomena and doppler effect in an AGN jet, which may be verified by the high-sensitivity observations in future.

\section*{Acknowledgements}
The first author (WPL) was supported by the National Natural Science Foundation of China under Grant U1231106 and the Scientific Research Foundation of Civil Aviation University of China under Grant 09QD15X. The author (LYL) acknowledges the support from the National Natural Science Foundation of China under Grant 11247274. The author (CCW) acknowledges the support from the Fundamental Research Funds for the Chinese Central Universities with Grant WK2030220004, as well as the National Natural Science Foundation of China under Grant 11073019.

\end{document}